\documentclass[11pt]{article}

\RequirePackage{amsthm,amsmath,amsfonts,amssymb}

\RequirePackage{enumerate}
\RequirePackage[colorlinks,citecolor=blue,urlcolor=blue]{hyperref}
\RequirePackage{graphicx}
\usepackage{algorithm}
\usepackage{algpseudocode}
\usepackage{natbib}
\usepackage{setspace}
\usepackage{geometry}

\theoremstyle{plain}

\theoremstyle{remark}

\title{Exact MCMC for Intractable Proposals}

\author{Dwija Kakkad \\ Department of Statistical Science \\ Duke University \\ {\tt dwija.kakkad@duke.edu} \and Dootika Vats \\ Department of Mathematics and Statistics \\ IIT Kanpur \\ {\tt dootika@iitk.ac.in}   }

\date{\today}
\begin{document}
\maketitle

\onehalfspacing

\begin{abstract}
Accept-reject based Markov chain Monte Carlo (MCMC) methods are the workhorse algorithm for Bayesian inference. These algorithms, like Metropolis-Hastings, require choosing a proposal distribution which is typically informed by the desired target distribution. Surprisingly, proposal distributions with unknown normalizing constants are not uncommon, even though for such a choice of a proposal, the Metropolis-Hastings acceptance ratio cannot be evaluated exactly. Across the literature, authors resort to approximation methods that yield inexact MCMC or develop specialized algorithms to combat this problem. We show how Bernoulli factory MCMC algorithms, originally proposed for doubly intractable target distributions, can quite naturally be adapted to yield an exact MCMC sampling method. We present three diverse and relevant examples demonstrating the usefulness of the Bernoulli factory approach to this problem.
\end{abstract}

\section{Introduction}
\label{sec:intro}

To sample from a distribution with density $\pi$, accept-reject based Markov chain Monte Carlo (MCMC) algorithms produce a $\pi$-ergodic Markov chain  using draws from a proposal distribution, $Q$. Algorithm~\ref{alg:AR_MCMC} details a typical accept-reject based MCMC algorithm, where the acceptance function $\alpha$ is appropriately chosen.  Let $q( \cdot |x)$ denote the density of $Q( \cdot|x)$ given the current state of the Markov chain, $x$. The popular  Metropolis-Hastings algorithm corresponds to  a particular choice of $\alpha$:
\begin{equation}
\label{eq:mh_ratio}
\alpha_{\text{MH}}(x, y) = \min \left\{1, \frac{\pi(y)q(x|y)}{\pi(x)q(y|x)}\right\}\,.
\end{equation}
If  $\pi(x) \propto \tilde{\pi}(x)$, where the normalizing constant is possibly unknown, it is still possible to implement Algorithm~\ref{alg:AR_MCMC} since the unknown normalizing constant cancels.  The wide use of the Metropolis-Hastings algorithm \citep{dunson2020hastings}  is a consequence of this very generality. 

Naturally, implementing any accept-reject MCMC algorithm requires the choice of a proposal distribution, $Q$. Different proposal distributions yield different algorithms like random walk Metropolis, Metropolis-adjusted Langevin algorithm \citep{roberts1996exponential}, Barker's proposal \citep{livingstone2022barker}, Hamiltonian Monte Carlo \citep{neal2011mcmc}, etc. Often, a proposal distribution is chosen  keeping in mind the geometry, support, and the general shape of the target density. Consequently, the literature has often employed a proposal density $q(\cdot|x)$ that although well-informed and appropriate, has an unknown normalizing constant.  A typical example is that of a truncated Gaussian proposal for target distributions with support on a set $\mathcal{A} \subset \mathbb{R}$. Here, for some $h > 0$,
\begin{equation}
    q(y | x) = \dfrac{\exp\left\{-\frac{(y-x)^2}{2h} \right\}  \mathbb{I}(y \in \mathcal{A})}{ \int_{\mathcal{A}}\exp\left\{-\frac{(s-x)^2}{2h} \right\} ds }\,.
\end{equation}
Technically, the normalizing constant is intractable, although very good approximations of the Gaussian cumulative distribution function are available across software. Even so, in the truest sense, employing truncated Gaussian proposals in a Metropolis-Hastings algorithm yields approximate Markov chains. In this work, we present a framework for exact MCMC for such intractable proposals, using Bernoulli factory methods. 
% It is natural to question the choice of such a proposal distribution at all. Surprisingly, as we will demonstrate, there are numerous instances of such proposal distributions being employed with adjustments made for the unavailability of the normalizing constant.

% We identify utility of Bernoulli factory MCMC algorithms for situations when $q$ is intractable, in that
% \begin{equation*}
%    q(y | x) = \dfrac{\tilde{q}(y|x)}{ r(x)}\,,
% \end{equation*}
% where $r(x)$ is the unknown normalizing function.  In such cases, the Metropolis-Hastings acceptance function reduces to
% \begin{equation*}
% \alpha_{\text{MH}}(x, y) = \min \left\{1, \frac{\pi(y) \tilde{q}(x|y)}{\pi(x) 
%  \tilde{q}(y|x)}   \dfrac{r(x)}{r(y)}\right\}\,,
% \end{equation*}
% and consequently cannot be evaluated. It is natural to question the choice of such a proposal distribution at all. Surprisingly, there are numerous instances of such proposal distributions being employed with adjustments made for the unavailability of $r(x)$.

% 
% 
\begin{algorithm}
\caption{Accept-reject based MCMC}\label{alg:AR_MCMC}
\begin{algorithmic}[1]
\State For $k = 1, 2, \dots, n$, let $X_k = x$ and repeat the following:
\State Draw $Y = y \sim Q( \cdot | x)$
\State Set $X_{k+1} = y$ with probability $\alpha(x,y)$ and $X_{k+1} = x$ with probability $1 - \alpha(x,y)$.
\end{algorithmic}
\end{algorithm}

In most MCMC applications, a typical way in which Step~3 of Algorithm~\ref{alg:AR_MCMC} is implemented is by drawing an independent $U \sim U(0,1)$ and setting $X_{k+1} = y$ if $U \leq \alpha(x,y)$ and $X_{k+1} = x$, otherwise. However, when $\pi$ is doubly intractable, in the sense that $\tilde{\pi}(x)$ itself cannot be evaluated, Algorithm~\ref{alg:AR_MCMC} is no longer employable. An alternate interpretation of Algorithm~\ref{alg:AR_MCMC} is presented in Algorithm~\ref{alg:alt_AR_MCMC}, where a Bern$(\alpha(x,y))$ event determines acceptance or rejection of the proposed $y$. Statistically, Algorithms~\ref{alg:AR_MCMC} and \ref{alg:alt_AR_MCMC} are equivalent.
Inspired by this representation in Algorithm~\ref{alg:alt_AR_MCMC}, \cite{herbei2014estimating,goncalves2017barkers,vats2021efficient} use Bernoulli factories to obtain a draw from Bern$(\alpha(x,y))$, when evaluating $\alpha(x,y)$ is not possible. For a function $l$ taking values in $[0,1]$, a Bernoulli factory is a mechanism of generating Bern$(l(\lambda))$ using Bern$(\lambda)$ draws, without evaluating $l$. Here $\alpha(x,y)$ represents $l(\cdot)$ and we will explain what $\lambda$ represents later. See \cite{keane1994bernoulli} for more information on Bernoulli factories outside the context of MCMC. Bernoulli factory MCMC algorithms have so far been used in situations when $\alpha(x,y)$ cannot be evaluated due to the double intractability of $\pi$. We highlight the use of Bernoulli factory MCMC algorithms for situations when $q$ is intractable.

\begin{algorithm}
\caption{Alternative representation of Algorithm~\ref{alg:AR_MCMC}}\label{alg:alt_AR_MCMC}
\begin{algorithmic}[1]
\State For $k = 1, 2, \dots, n$, let $X_k = x$ and repeat the following:
\State Draw $Y = y \sim Q( \cdot | x)$
\State Draw $B \sim \text{Bern} \left(\alpha(x,y) \right)$ independent of $Y$
\State Set $X_{k+1} = y$ if $B = 1$ and $X_{k+1} = x$ if $B = 0$.
\end{algorithmic}
\end{algorithm}

A Bernoulli factory is not possible for all functions $l$, and in fact it has been shown that there cannot exist a Bernoulli factory to generate events of probability $\alpha_{\text{MH}}(x,y)$\citep{latuszynski2011simulating}. Nonetheless, the Metropolis-Hastings acceptance function is only one choice of an acceptance function. There are many other acceptance functions that can yield a $\pi$-stationary Markov chain. 
% In fact, a valid acceptance function is one that satisfies
% \begin{equation*}
%     \alpha(x,y) q(y | x) \pi(x) = \alpha(y,x) q(x | y) \pi(y)\,.
% \end{equation*}
For instance, the Barker's acceptance function \citep{barker1965monte} below is a common alternative to the Metropolis-Hastings acceptance function:
\begin{equation}
    \label{eq:barker_acc}
\alpha_{\text{B}}(x, y) = \frac{\pi(y)q(x|y)}{\pi(y)q(x|y) + \pi(x)q(y|x)}\,.
\end{equation}
See \cite{agrawal2023optimal} for a non-exhaustive list of other acceptance functions. The popularity of Metropolis-Hastings over other acceptance functions is due to the result of \cite{peskun1973optimum}, proving that the Metropolis-Hastings acceptance function is optimal, in the sense that the asymptotic variance of ergodic averages is always smaller due to Metropolis-Hastings, compared to any other acceptance function. However, \cite{latuszynski2013clts} showed that the Barker's acceptance function at worst, doubles the asymptotic variance of Monte Carlo averages. Motivated by this, \cite{goncalves2017barkers} showed that unlike $\alpha_{\text{MH}}(x,y)$, a Bernoulli factory can be constructed to generate events of probability $\alpha_{\text{B}}(x,y)$, when $\pi$ is doubly intractable.

To efficiently implement the Bernoulli factory algorithm of \cite{goncalves2017barkers}, one requires a tight upper bound of $\pi(y) q(x | y)$. When $\pi$ is doubly intractable, finding tight bounds for it can be challenging since $\pi$ typically arises from a complicated Bayesian model. This requirement has restricted the use of Bernoulli factory MCMC for general doubly intractable targets. However, when $q$ is intractable, it is still such that it must be possible to sample from $Q$, and thus, often tight bounds are available, making it feasible to employ exact MCMC for intractable proposals via Bernoulli factories. 

We demonstrate the utility of our proposed strategy via three examples\footnote{Reproducible \texttt{R} code is available at: https://github.com/dwija04/BernoulliFactoryProposals.}. We continue the simple one-dimensional truncated Gaussian proposal example as a way to demonstrate our proposed methods and a proof of concept. Our second example is the repelling-attracting Metropolis (RAM) algorithm of \cite{Tak_2018}, where the proposal distribution is intractable by design, and we compare our Bernoulli factory solution with an auxiliary variable solution employed by the authors. 
% Our Bernoulli factory solution can allow for a wider class of RAM proposal distributions to be employed, whereas the authors' auxiliary variable solution is restricted to symmetric proposals. 
Our final example is that of the estimation of a Cox process via constrained modulated Gaussian process developed by \cite{pmlr-lopez-gauss-cox}. Intractability of the proposal in the original work was handled by implementing approximate Metropolis-Hastings. Our Bernoulli factory solution is both faster and exact. In all three examples, tight bounds on the normalizing constants are easily available, and the Bernoulli factories are fairly efficient.

\section{Bernoulli Factories for MCMC}
\label{sub:two-coin-target}

% \subsection{Two-coin algorithm for doubly intractable targets}
We first provide a short review of the two-coin Bernoulli factory proposed by \cite{goncalves2017barkers} to implement Barker's algorithm for doubly intractable targets. Recall that $\pi(x)$ is such that $\pi(x) \propto \tilde{\pi}(x)$ where further assume that
\[
\tilde{\pi}(x)  = \int g(x, \eta) d\eta\,,
\]
with $\eta$ being a nuisance parameter. Such a $\pi$ is an instance of a doubly intractable posterior. \cite{vats2021efficient} discuss that such target densities can be seen in Bayesian diffusion models, Bayesian models over constrained spaces, and Bayesian generalized linear models. The goal remains to run a $\pi$-ergodic MCMC algorithm, but evaluating $\alpha_{\text{MH}}$ is no longer possible. The two-coin algorithm of  \cite{goncalves2017barkers} is a Bernoulli factory that, given $x$ and $y$, generates Bern$(\alpha_{\text{B}}(x,y) )$ events without evaluating $\alpha_{\text{B}}(x,y)$. This two-coin algorithm is then used in Step~3 of Algorithm~\ref{alg:alt_AR_MCMC} to yield a $\pi$-invariant Markov chain. 

Let $x$ be the current state of the Markov chain and let $y$ be the proposed state. The two-coin algorithm builds on the following construction.  Suppose for some $0 \leq p_x, p_y \leq 1$ and $c_x, c_y > 0$, we can write
\[
\pi(x)q(y|x) = c_xp_x  \quad \text{and} \quad \pi(y)q(x|y) = c_yp_y.
\]
Consequently, we can rewrite $\alpha_{\text{B}}(x,y)$ as 
\[
\alpha_{\text{B}}(x, y) = \frac{c_yp_y}{c_xp_x + c_yp_y}\,.
\]
A restriction here is that $c_x$ and $c_y$ must be such that they can be evaluated. One way to arrive at $c_x$ and $p_x$ is to find a $c_x$ satisfying 
\begin{equation}
\label{eq:cx_bound}
    \pi(x)q(y|x) \leq c_x \Rightarrow p_x = \dfrac{\pi(x)q(y|x)}{c_x}\,.
\end{equation}
A similar arrangement yields $c_y$ and $p_y$. The two-coin algorithm of \cite{goncalves2017barkers} presented in Algorithm~\ref{alg:twocoin} is a Bernoulli factory that outputs 1 with probability  $\alpha_{\text{B}}(x, y)$. 
\begin{algorithm}
\caption{Two-coin algorithm}\label{alg:twocoin}
\begin{algorithmic}[1]
\State Draw $C_1 \sim $ Bern$\left(\frac{c_y}{c_x + c_y} \right)$
\If{$C_1 = 1$}
    \State Draw $C_2 \sim$ Bern$(p_y)$
        \If{$C_2 = 1$}  output $1$
        % \State 
        \Else ~go to Step 1
\EndIf
\EndIf  
\If{$C_1 = 0$}
    \State Draw $C_2 \sim$ Bern$(p_x)$
    \If{$C_2 = 1$}  output $0$
    \Else ~go to Step 1
    \EndIf
\EndIf
\end{algorithmic}
\end{algorithm}

In Algorithm~\ref{alg:twocoin}, Steps~5 and 11 loop back to Step~1, making it possible for the algorithm to be stuck in long loops. In fact, when the bounds $c_y$ and $c_x$ are too large, $p_y$ and $p_x$ are too small so that both Bern$(p_y)$ and Bern$(p_x)$ return failures, forcing a loop. More specifically, \cite{goncalves2017barkers} explain that the expected number of loops needed until the algorithm returns a realization are
\[
\frac{c_x + c_y}{c_xp_x + c_yp_y} = \frac{c_x + c_y}{\pi(x)q(y|x) + \pi(y)q(x|y)}\,.
\]
Notably, the larger the bounds $c_x$ and $c_y$, the more inefficient the two-coin algorithm will be. It is fairly challenging to obtain tight bounds $c_x$ and $c_y$ in a typical Bayesian problem, as this often involves bounding the likelihood, which scales exponentially with data size. \cite{vats2021efficient} proposed a small modification to the two-coin algorithm to arrive at a portkey two-coin algorithm that can be significantly more efficient, however the acceptance probability is smaller than $\alpha_{\text{B}}(x, y)$, although still $\pi$-invariant.

\section{Exact MCMC for intractable proposals}
% \subsection{Two-coin algorithm for intractable proposals}
\label{sec:two-coin-proposal}

Consider an intractable proposal density of the form $q(y|x) \propto \tilde{q}(y|x)$, 
where the proportionality constant is unknown. As discussed in Section~\ref{sec:intro}, in this case as well, $\alpha_{\text{MH}}$ cannot be evaluated, unless of course, $q(y|x)$ is independent of $x$.  As we will demonstrate in Section~\ref{sec:examples}, it is not uncommon for algorithms to employ such proposal distributions. As it turns out, in order to implement the two-coin Bernoulli factory algorithm, one just needs to obtain the bound in \eqref{eq:cx_bound}. For doubly intractable posteriors, the task reduces to bounding $\pi$ and for intractable proposals, it reduces to bounding $q$.

% Let $Q( \cdot |x)$ be a proposal distribution with density $q(\cdot | x)$ defined on a support $\mathcal{A}$. The proposal $Q$ is such that it is possible to obtain realizations from it, but $q$ is intractable in that
Consider a proposal distribution $Q(\cdot|x)$ defined on $\mathcal{A} \subseteq \mathcal{Y}$ with density
\[
q(y|x) = \dfrac{\tilde{q}(y|x)}{r(x)}\,,
\]
where $\tilde{q}(y|x)$ is known, but the normalizing function $r(x)$ is unknown and of the form
\begin{equation}
    \label{eq:rofx}
    r(x) = \int_{\mathcal{A}} \tilde{q}(s | x) ds \,.
\end{equation}
The Barker's acceptance function, in this case can be written as
\begin{equation}
\label{eq:barker_simp}
    \alpha_{\text{B}}(x,y) = \dfrac{\pi(y) \tilde{q}(x|y) r(x)}{\pi(y) \tilde{q}(x|y) r(x) + \pi(x) \tilde{q}(y|x) r(y)} \,.
\end{equation}
To implement the two-coin algorithm, we have to upper bound the numerator and the denominator terms as in \eqref{eq:cx_bound}. However, since $\pi$ and $\tilde{q}$ can be evaluated, we just need to upper bound $r(\cdot)$ to obtain a usable upper bound. Specifically, let $r(y) \leq b_y$, then set
\begin{align}
\label{eq:cx} 
    c_x &:= \pi(x) \tilde{q}(y|x) b_y  \quad \text{and} \quad p_x := \dfrac{r(y)}{b_y} \,.
\end{align}
We use subscript $x$ in $p_x$ to reinforce that in Algorithm~\ref{alg:twocoin}, a Bern$(p_x)$ success yields the next iterate to be $x$. A similar construction can be done for the numerator term, so that $r(x) \leq b_x$ and
\begin{align}
    c_y &:= \pi(y) \tilde{q}(x|y) b_x  \quad \text{and} \quad p_y := \dfrac{r(x)}{b_x} \,.\label{eq:cy}
\end{align}
With these choices of $c_x, c_y, p_x, p_y$, Algorithm~\ref{alg:twocoin} can be implemented directly. 

The last piece of the puzzle is to arrive at a way of simulating Bern$(p_y)$ events when $p_y$ cannot be evaluated. Given the form of $r(x)$ in \eqref{eq:rofx}, this can often be done through  importance sampling. Specifically, consider a distribution $F$ defined on a support that contains $\mathcal{A}$ with density $f$ such that for all $m \in \mathcal{A}$,
\[
    \dfrac{\tilde{q}(m | x)}{f(m) b_x}  \leq 1\,.
\]
Then,
\begin{enumerate}[(i)]
    \item Draw $M \sim F$.
    \item Draw $C_2 | M \sim \text{Bern} \left( \dfrac{\tilde{q}(M | x)}{f(M) b_x} \right)$.
    \item Marginally, $C_2 \sim \text{Bern}(p_y)$\,.
\end{enumerate}
Thus, Steps~3 and 9 of Algorithm~\ref{alg:twocoin} can be implemented quite simply. A single draw from $F$ suffices, and the resulting Markov chain is exact. As will be seen in the examples in Section~\ref{sec:examples}, finding such an $F$ is straightforward for truncated proposals.
% \subsection{Exact Barker's algorithm for intractable proposals}
% \label{sub:final_algo}
 
We now provide the final algorithm for implementing Barker's accept-reject algorithm via Bernoulli factories for intractable proposals. To implement the algorithm, one requires bounds for the normalizing functions of the proposal, $r(x)$ and $r(y)$, denoted here by $b_x$ and $b_y$, respectively.
\begin{algorithm}
\caption{Bernoulli factory Barker's algorithm}\label{alg:final}
\begin{algorithmic}[1]
\State \textbf{Input:} Current state $X_k = x$
\State \textbf{Output:} Next state $X_{k+1}$
\State Draw $y \sim Q(\cdot | x)$ 
\State Obtain expressions for $c_x$, $p_x$, $c_y$, and $p_y$ as given in \eqref{eq:cx} and \eqref{eq:cy}
\State Let $B$ be the output from Algorithm \ref{alg:twocoin}
\If{$B = 1$}
    \State \textbf{return} $X_{k+1} = y$
\Else 
    \State \textbf{return} $X_{k+1} = x$
\EndIf
\end{algorithmic}
\end{algorithm}
Algorithm~\ref{alg:final} implements an exact $\pi$-invariant Markov chain using Barker's accept-reject function. Typically, any proposal $Q$ has a tuning parameter which is chosen according to optimal tuning strategies available for Metropolis-Hastings acceptance functions \citep{roberts2001optimal}. For random walk proposals, Metropolis-Hastings acceptance function desires a roughly $44\%$ acceptance for one-dimension problems and $23\%$ acceptance in high dimensions. Recently, for Barker's acceptance functions, \cite{agrawal2023optimal} show that an optimal strategy for random walk proposals is to tune the proposal to $25\%$ acceptance in one dimension and $16\%$ acceptance in high-dimensions. Further, \cite{agrawal2023optimal} show that for a given random walk proposal $Q$, tuned to their own optimal strategies, Metropolis-Hastings accept-reject and Barker's accept-reject yield a relative efficiency of 1.39. This agrees with \cite{peskun1973optimum}, that Metropolis-Hastings is more efficient, but provides a clearer picture than \cite{latuszynski2013clts} who showed that Metropolis-Hastings has an efficiency no better than twice as that of Barker's. 
% Although, the proposals we employ are not random walk proposals, we find these recommended strategies are effective in our implementations.

\section{Examples}
\label{sec:examples}

\subsection{Truncated Gaussian proposals}

\begin{figure*}
\centering
\includegraphics[width=0.45\columnwidth]{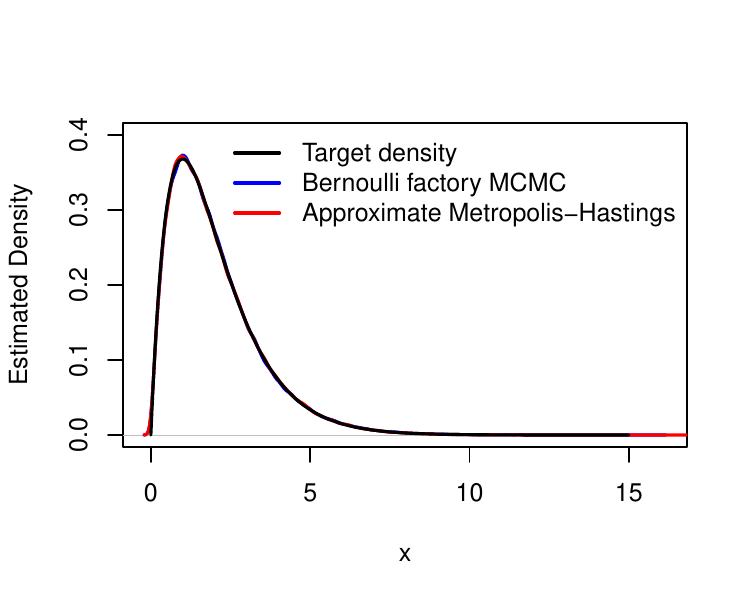} \hspace{.8cm}
\includegraphics[width=0.45\columnwidth]{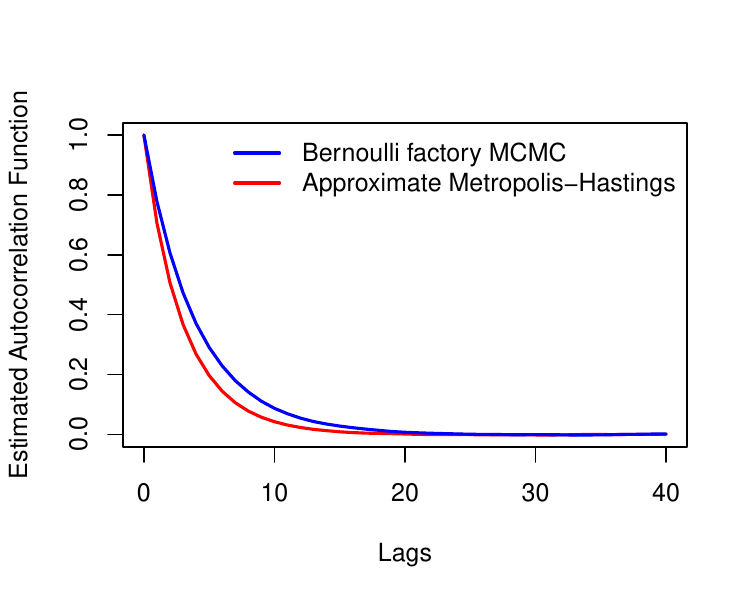}
\caption{Gamma target: (Left) Density plot for one run of Metropolis-Hastings and Barker's algorithms and (right) estimated autocorrelation function plots for Metropolis-Hastings and Barker's algorithms.}
\label{fig:density-exact}
\end{figure*}

A Gaussian distribution is a popular choice for a proposal distribution. However, for target distributions defined on a restricted support, a Gaussian proposal can be inefficient, as values proposed outside the support are immediately rejected. A solution often employed is to use a truncated Gaussian proposal, truncated to the support of the target distribution \citep{chen2024coreset,sarkar2024mr}. For a target, $\pi$, with support $\mathcal{A} \subset \mathbb{R}$, consider a Gaussian proposal with mean $x$, variance $h$, truncated to the set $\mathcal{A}$:
\begin{equation}
    q(y | x) = \dfrac{ \dfrac{1}{\sqrt{2\pi h}}\exp\left\{-\dfrac{(y-x)^2}{2h} \right\}  \mathbb{I}(y \in \mathcal{A})}{ \displaystyle\int_{\mathcal{A}}  \dfrac{1}{\sqrt{2\pi h}}\exp\left\{-\dfrac{(s-x)^2}{2h} \right\} ds }\,.
\end{equation}
As described in Section~\ref{sec:intro}, although excellent approximations of the Gaussian cumulative distribution functions are available, the resulting Metropolis-Hastings algorithm remain theoretically inexact. Alternatively, it is actually quite straightforward to implement the Barker's algorithm via the two-coin Bernoulli factory. See that in this case,
\begin{equation*}
    \dfrac{c_y}{c_y + c_x} = \dfrac{\pi(y) b_x}{\pi(y) b_x + \pi(x) b_y}\,,
\end{equation*}
where,
\begin{align*}
r(x) &= \displaystyle \int_{\mathcal{A}} \dfrac{1}{\sqrt{2\pi h}} \exp\left\{-\dfrac{(s-x)^2}{2h} \right\} ds   \leq 1 =: b_x\,, \\ 
r(y) & = \displaystyle \int_{\mathcal{A}} \dfrac{1}{\sqrt{2\pi h}} \exp\left\{-\dfrac{(s-y)^2}{2h} \right\} ds   \leq 1 =: b_y\,. 
\end{align*}
A simple upper bound is thus available on $r(x)$. To simulate Bern$(p_y)$ coins, we set $F = N(x,h)$, so that
\[
    \dfrac{\tilde{q}(m | x)}{f(m) b_x}  = \mathbb{I}( M \in \mathcal{A}) \leq 1\,.
\]
Consequently,
\begin{enumerate}[(i)]
    \item Draw $M \sim N(x, h)$.
    \item Set $C_2|M =  \mathbb{I}( M \in \mathcal{A})$\,. Then $C_2 \sim \text{Bern}(p_y)$.
\end{enumerate}
Note that in case of truncated proposals, $p_y$ is either 0 or 1, which means that we can directly set the next iterate to be $y$ if $p_y = 1$ without performing a Bernoulli draw. We now have all the ingredients to implement the two-coin algorithm for a given target density. As a toy example, consider the Gamma$(\alpha, \beta)$ target distribution  so that
\[
\pi(x) \propto x^{\alpha - 1} e^{-\beta x} \qquad x > 0\,.
\]
The support is $\mathcal{A} = \{x \in \mathbb{R}: x > 0\}$. We set $\alpha = 2$ and $\beta = 1$ and tune $h$ according to the optimal acceptance rates known for Metropolis-Hastings and Barker's. For a visual comparison, we first run both the approximate Metropolis-Hastings algorithm (using the \texttt{pnorm} function in \texttt{R} to approximate the distribution function of a Gaussian) and the exact Bernoulli factory Barker's algorithm for $10^6$ iterations and compare the estimated density and estimated autocorrelation function plots. Results are in Figure~\ref{fig:density-exact} where both estimated density plots are similar and as expected, the Barker's algorithm demonstrates slightly more persistent autocorrelation than the Metropolis-Hastings algorithm. The \texttt{pnorm} function in \texttt{R} uses the approximation by \cite{cody1969rational} that is accurate up to 18 decimal digits, and thus, we do not expect a visual difference in the estimated density plots. Such approximations are not available for a multivariate truncated Gaussian distribution, which we will focus on in Section~\ref{sub:cox}.

To assess the computational burden added by the Bernoulli factory, we run 100 Markov chains of length $10^6$ each to obtain the average loops in the Bernoulli factory, and calculate the average of the maximum loops taken in each run. The results are provided in Table~\ref{tab:bern_loops_trunc}. The two-coin algorithm is quite efficient, requiring an average of 1.33 loops to generate an accept-reject coin. The average of the maximum number of loops required in runs of length $10^6$ was 16.  For comparison, we present effective sample size (ESS), computational time, and effective sample size per second in Table~\ref{tab:trunc_ess} for both Metropolis-Hastings and Barker's algorithms. Noticeably, the computational time for both algorithms is similar. The effective sample size from the Metropolis-Hastings implementation is roughly 1.3 times the effective sample size from Barker's implementation, in line with the theory discussed in Section~\ref{sec:two-coin-proposal}.

\begin{table}
\centering
% \small 
% \tabcolsep=2pt % Reduce space between columns
\caption{Gamma target: Average loops of the two-coin algorithm}
\label{tab:bern_loops_trunc}
\begin{tabular}{lc}
\hline
\textbf{Loops} & \textbf{Bernoulli factory}  \\
\hline
Mean loops  &  $1.33$ \\
Maximum loops & $16.00$ \\
\hline
\end{tabular}
\end{table}

\begin{table}
\centering
% \small % Reduce font size if desired
% \tabcolsep=2pt % Reduce space between columns
\caption{Gamma target: Effective sample size}
\label{tab:trunc_ess}
\begin{tabular}{lcc} 
\hline
\textbf{Proposal} & \textbf{Bernoulli} & \textbf{Metropolis} \\
    & \textbf{Factory} & \textbf{Hastings} \\
\hline
ESS & 120682 & 160222 \\
ESS per second & 4221 & 6353\\ 
Avg computing time (s) & 30.64 & 27.63\\
\hline
\end{tabular}
\end{table}

\subsection{Repelling-Attracting Metropolis algorithm}

The RAM algorithm of \cite{Tak_2018} was developed to encourage mode-jumping in multi-modal distributions. The key idea behind the RAM algorithm is to propose the next move in two steps:  a forced downward step that intentionally looks to move in areas of low probability and a subsequent forced upwards move. This helps the Markov chain jump modes more effectively than standard random-walk proposals.

The RAM proposal is constructed in the following way. Let $\pi$ be the target density and let $s(\cdot|x)$ be a symmetric proposal density. Their down-up proposal density is denoted by $q^{\text{DU}}(y | x)$ and a draw from this is obtained in the following way for a small $ \epsilon > 0$:
\begin{enumerate}
    \item Propose $x'$ from $s(\cdot |x)$ and accept with probability 
    \[
    \alpha_{\epsilon}^{\text{D}}(x, x') = \min \left\{ 1, \frac{\pi(x) + \epsilon}{\pi(x') + \epsilon}\right\}.
    \] 
    That is, continue proposing from $s$ until a draw is accepted. This forces a downward move, and the density of the accepted $x'$ is
    \[
    q^\text{D}(x'|x) = \frac{s(x'|x)\alpha_{\epsilon}^{\text{D}}(x, x')}{A^{\text{D}}(x)}\,, 
    \]
    where $A^\text{D}(x)$ is the normalizing constant
    \begin{equation}
    \label{ram:AD}
        A^\text{D}(x) = \int s(x'|x)\alpha_{\epsilon}^\text{D}(x, x')dx'\,.
    \end{equation}

    \item Propose $y$ from $s(\cdot |x')$ and accept with probability 
    \[
    \alpha_{\epsilon}^\text{U}(x', y) = \min \left\{ 1, \frac{\pi(y) + \epsilon}{\pi(x') + \epsilon}\right\}\,.
    \] 
    That is, continue proposing from $s$ until a draw is accepted. This forces an upward move, and the density of the accepted $y$ is 
    \[
    q^{\text{U}}(y|x') = \frac{s(y|x')\alpha_{\epsilon}^{\text{U}}(x', y)}{A^{\text{U}}(x')}, 
    \] 
    where $A^{\text{U}}(x')$ is the normalizing constant  $A^{\text{U}}(x') = \int s(y|x')\alpha_{\epsilon}^{\text{U}}(x', y)dy.$
\end{enumerate}
The purpose of $\epsilon > 0 $ is to avoid divide-by-zero errors. Given $x$, the final proposal density is obtained by integrating out the intermediate downward move
\[
q^{\text{DU}}(y|x) = \int q^{\text{D}}(x'|x)q^{\text{U}}(y|x') dx'\,.
\]
For most choices of $s$, $q^{\text{DU}}$ is intractable. Employing this proposal density, and critically due to the fact that $s$ is a symmetric proposal density, it can be shown that the Metropolis-Hastings ratio reduces to
\begin{equation}
    \label{eq:RAM_ratio}
    \dfrac{\pi(y)q^{\text{DU}}(x|y)}{\pi(x)q^{\text{DU}}(y|x)} = \dfrac{\pi(y)A^{\text{D}}(x)}{\pi(x)A^{\text{D}}(y)}\,.
\end{equation}
Since $A^{\text{D}}$ is intractable, the Metropolis-Hastings ratio cannot be evaluated and thus implementing the standard Metropolis-Hastings algorithm is not possible.

\subsubsection{The auxiliary variable approach}

\cite{Tak_2018} employ an auxiliary variable approach to generate a Markov chain using $q^{\text{DU}}$ as a proposal. The method originally introduced in \cite{auxiliary}, involves augmenting the Markov chain state-space with an auxiliary variable, $z$, in such a way that the marginal density for $x$ remains  $\pi$, but the Metropolis-Hastings acceptance ratio for the joint chain is tractable. Algorithm~\ref{alg:aux_ram} provides the detailed steps. After generating $y$ from $q^{\text{DU}}$, generate $z^*$ given $y$ using the downward density $q^{\text{D}}$. Consequently, the joint target density is $\pi(x, z) = \pi(x)s(z|x)$, but employing accept-reject for the proposed $(y, z^*)$ becomes tractable. The method requires additional computational effort due to the sampling of $z^*$.

% \begin{remark}
%     Critically, the auxiliary variable approach is only applicable when $s$ is a symmetric proposal. In \cite{Tak_2018}, the authors mention that it would be useful to employ an anti-Langevin proposal in $s$, but such a choice does not allow for exact sampling via the auxiliary variable approach.
% \end{remark}

\begin{algorithm}
\caption{RAM algorithm using auxiliary variables}
\label{alg:aux_ram}
\begin{algorithmic}[1]
\State Set initial values $X_0, Z_0$
\State \textbf{for} $t = 1, 2, \dots$, set $(x, z) = (X_{t}, Z_{t})$ and
    \State Repeatedly sample $x' \sim s(\cdot | x)$ and $U_1 \sim \text{Uniform}(0, 1)$ until $U_1 < \min\left\{ 1, \frac{\pi(x) + \epsilon}{\pi(x') + \epsilon}\right\}$

    \State Repeatedly sample $y \sim s(\cdot | x')$ and $U_2 \sim \text{Uniform}(0, 1)$ until $U_2 < \min\left\{ 1, \frac{\pi(y) + \epsilon}{\pi(x') + \epsilon}\right\}$

    \State Repeatedly sample $z^* \sim s(\cdot | y)$ and $U_3 \sim \text{Uniform}(0, 1)$ until $U_3 < \min\left\{ 1, \frac{\pi(y) + \epsilon}{\pi(z^*) + \epsilon}\right\}$

    \If{for $U_4 \sim \text{Uniform}(0,1)$, $U_4 < \min \left\{1 , \frac{\pi(y)\min\left\{ 1, \frac{\pi(x) + \epsilon}{\pi(z) + \epsilon}\right\}}{\pi(x)\min\left\{ 1, \frac{\pi(y) + \epsilon}{\pi(z^*) + \epsilon}\right\}} \right\}$}
        \State Set $(X_{t+1}, Z_{t+1}) = (y, z^*).$
    \Else
        \State Set $(X_{t+1}, Z_{t+1}) = (x, z).$
    \EndIf
\end{algorithmic}
\end{algorithm}

\subsubsection{RAM algorithm using Bernoulli factories}

We implement the two-coin Barker's algorithm for the RAM proposal. The Barker's acceptance function in this case simplifies to
\begin{equation}
\label{eq:barker-ram}
    \alpha_{\text{B}}(x, y) = \frac{\pi(y)A^{\text{D}}(x)}{\pi(x)A^{\text{D}}(y) + \pi(y)A^{\text{D}}(x)}\,.
\end{equation}
Comparing \eqref{eq:barker-ram} with \eqref{eq:barker_simp}, we recognize that
\begin{align*}
 r(x) & = A^{\text{D}}(x) \\ 
 &= \int s(x'|x)\alpha_{\epsilon}^{\text{D}}(x, x')dx' \\
 &\leq \int s(x'|x)dx' \\
 &\leq 1 =: b_x\,.
\end{align*}
Similarly,
\[
r(y) = \int s(x'|y)\alpha_{\epsilon}^{\text{D}}(y, x')dx' \leq 1 =: b_y\,.
\]
Here again, simple and effective upper bounds are available on the unknown probabilities. In order to simulate $\text{Bern}(p_y = r(x)/b_x)$ coins, we do the following, setting $f(\cdot) = s(\cdot|x)$:
\begin{enumerate}
    \item Draw $M \sim s(\cdot|x)$
    \item Draw $C_2|M \sim \text{Bern}(\alpha_{\epsilon}^{\text{D}}(x, M) )$
    \item Marginally $C_2 \sim \text{Bern}(p_y)$.
\end{enumerate}
% With this, the ingredients for implementing the two-coin algorithms are complete.

% \begin{remark}
% Unlike the auxiliary variable approach, the Bernoulli factory implementation is not restricted to symmetric choices of $s$. 
% \end{remark}

\subsubsection{Sensor network localization}
\begin{figure*}
\centering
\includegraphics[width=0.45\columnwidth]{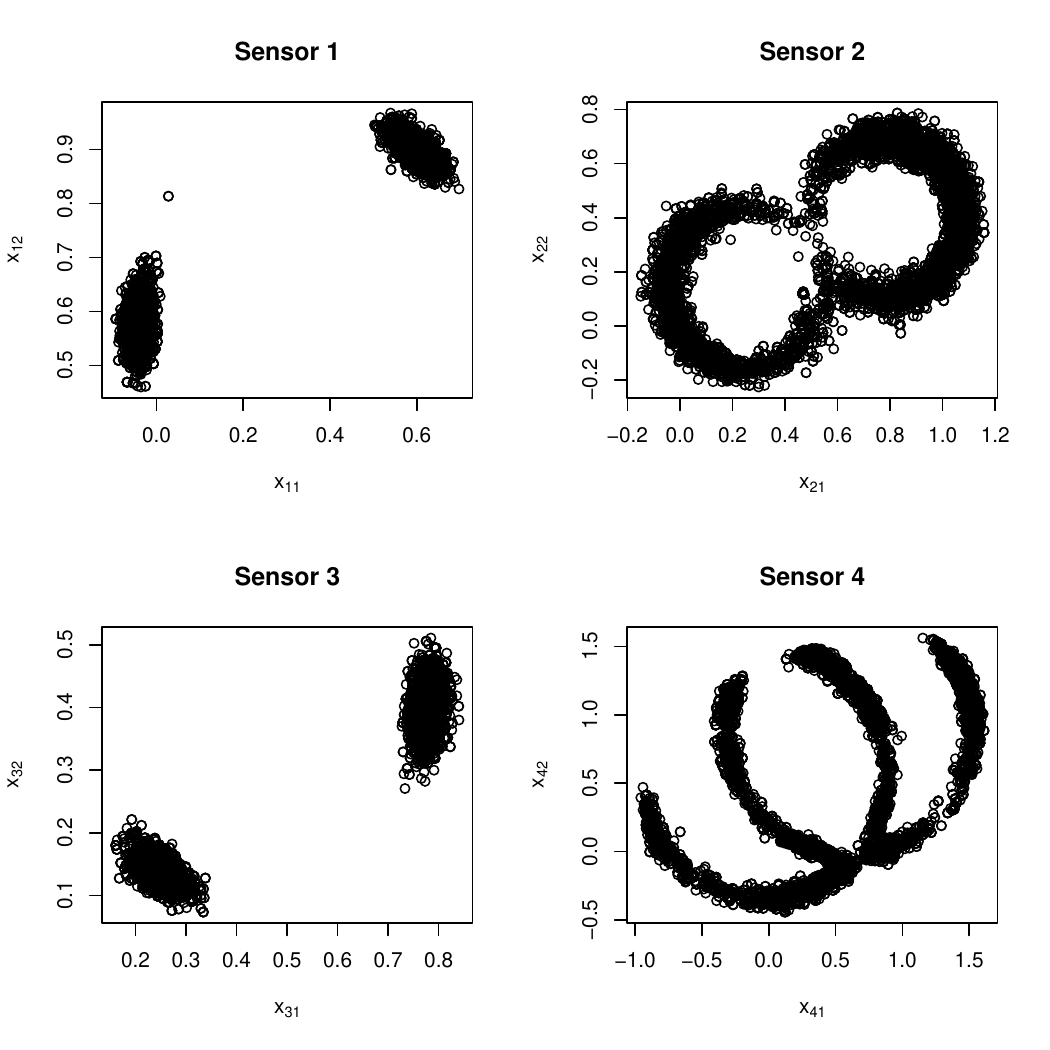} 
% \hspace{1cm}
\includegraphics[width=0.45\columnwidth]{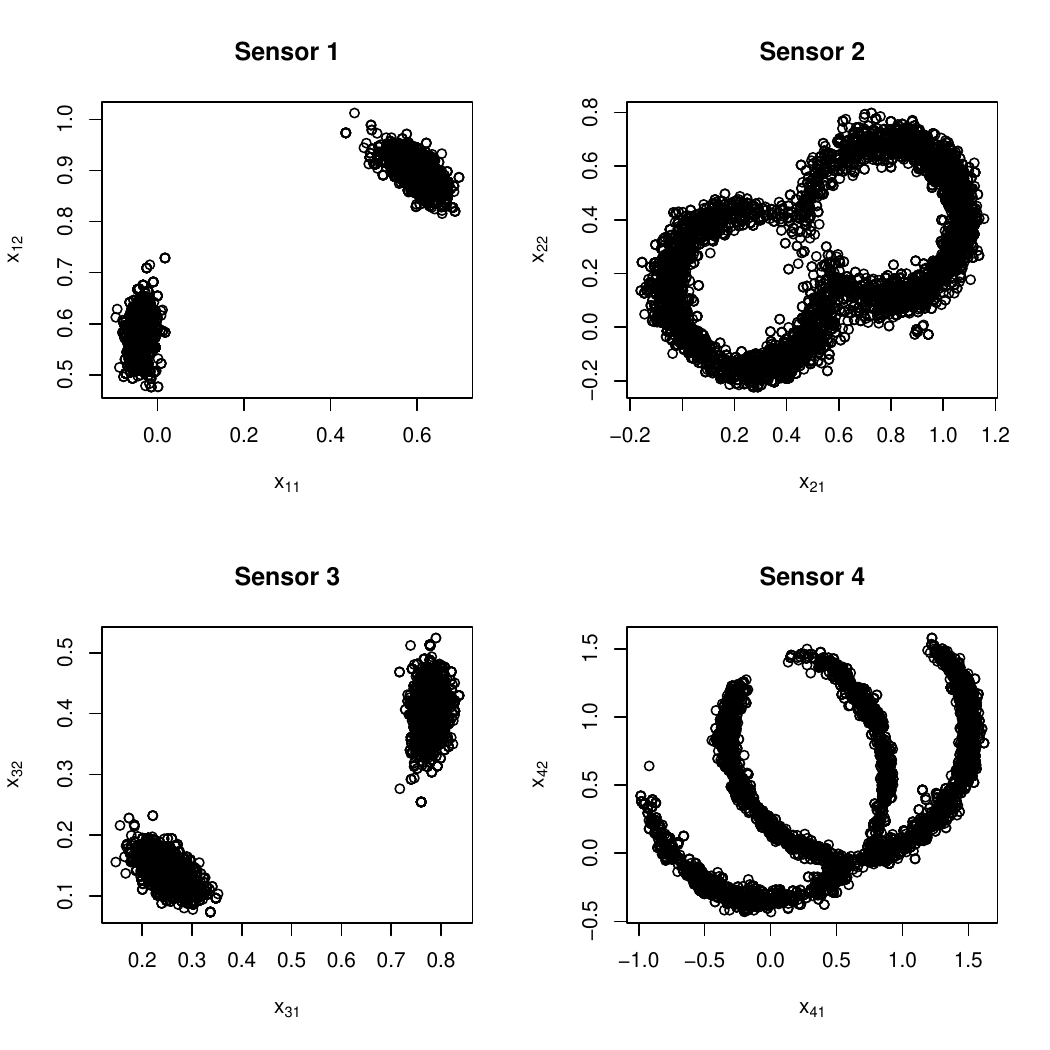}
\caption{Sensor network localization: Scatterplots using auxiliary variable method (left) and Barker's algorithm (right).}
\label{fig:sensors-both}
\end{figure*}

Consider a sensor network localization example from \cite{Ihler} where the goal is to search for unknown sensor locations using noisy distance data. There are 6 sensors scattered on a two-dimensional plane and the location of 4 of them is unknown. Let $x^\top_k = (x_{k1}, x_{k2})$ denote the co-ordinates of the $k^{\text{th}}$ sensor, and $y_{ij}$ denote the Euclidean distance between the $i^{\text{th}}$ and $j^{\text{th}}$ sensors. The variable $w_{ij}$ is 1 if the distance between $x_i$ and $x_j$ is observed and zero otherwise. The complete model is
$$
w_{ij} | x_1, \dots x_4 \overset{\text{ind}}{\sim} \text{Bern}\left(\exp\left(-\frac{\lVert x_i - x_j\rVert^2}{2 \times 0.3^2}\right)\right)\,,
$$
and
$$
y_{ij} | (w_{ij} = 1), x_1, \dots x_4 \overset{\text{ind}}{\sim} 1(\lVert x_i - x_j \rVert, 0.02^2)\,.
$$
We borrow the simulation setup of \cite{Tak_2018}. For each unknown location, assume a diffuse bivariate Gaussian prior with mean $(0, 0)^{\top}$ and covariance matrix $10^2 \times I_2$. The 8-dimensional posterior distribution of $(x_1, x_2, x_3, x_4)$ is highly multimodal. \cite{Tak_2018} employed a Metropolis-within-Gibbs approach and use the RAM proposal for the update of each of the four locations, yielding four accept-reject steps, one for each location. For each location, the proposal distribution, $s$, is a random-walk bivariate normal with covariance matrix  $1.08 I_2$. Coding implementations of their RAM algorithm are available via their official code submission. For equal comparison, we employ the same scheme, replacing the auxiliary variable Metropolis-Hastings scheme with the two-coin Barker's scheme. We first implement both auxiliary variable and Bernoulli factory chains for a single run for length $10^6$ and a thinned plot of the samples is in  Figure~\ref{fig:sensors-both}. Both methods seem to have jumped modes reasonably well. 

In Tables~\ref{tab:RAM_loops} and \ref{tab:RAM_multiess_table}, we quantitatively compare the performance of the two methods by replicating chains of length $2 \times 10^5$, 100 times. First, we note that the Bernoulli factory and auxiliary variable computational burdens are quite similar, with the auxiliary variable sampler being marginally slower. The effective sample size is estimated using the methods of \cite{vats2019multivariate} implemented in the \texttt{R} package \texttt{mcmcse} \citep{mcmcse}. As expected, Barker's algorithm is slightly less statistically efficient.
 % however the availability of choosing asymmetric $s$ in Bernoulli factory implementations will provide a lot more flexibility to users.
\begin{table}
\centering
% \small 
% \tabcolsep=6pt
\caption{RAM: Comparison of average and maximum loops}
\label{tab:RAM_loops}
\begin{tabular}{lcccc}
\hline
\textbf{Loops}  & \multicolumn{2}{c}{\textbf{Average}} & \multicolumn{2}{c}{\textbf{Maximum}} \\
\hline
 & Bernoulli & Auxiliary & Bernoulli & Auxiliary \\
       & factory & variable & factory & variable \\
\hline
Sensor 1 & 1.07 & 1.07 & 6.70 & 6.54 \\
Sensor 2 & 1.08 & 1.08 & 6.96 & 6.86 \\
Sensor 3 & 1.05 & 1.05 & 6.04 & 5.85 \\
Sensor 4 & 1.13 & 1.13 & 10.75 & 10.91 \\
\hline
\end{tabular}
\end{table}

\begin{table}
\centering
% \small
% \tabcolsep=2pt
\caption{RAM: ESS and computational time}
\label{tab:RAM_multiess_table}
\begin{tabular}{lcc}
\hline
\textbf{} & \textbf{Bernoulli} & \textbf{Auxiliary}\\
    & \textbf{factory} & \textbf{variable} \\
\hline
ESS &  164  & 201 \\
ESS$/$s  & 0.0387 & 0.0428 \\
Time (s) & 4442.178 &  4905.473 \\
\hline
\end{tabular}
\end{table}

\subsection{Gaussian process modulated Cox process}
\label{sub:cox}

% Consider an inhomogenous Poisson point process $X$ with non negative density $\lambda(\cdot)$, where $X$ is a countable subset of $\mathcal{S} \subset \mathbf{R}^d$. Let $N \in \mathbb{N}$ be a random variable denoting the number of points in  $X$. Let $X_1, X_2, \dots X_n$ be a set of $n$ i.i.d. random variables on $S$.
A Cox process is an extension of an inhomogeneous Poisson process where the intensity $\lambda(\cdot)$ is sampled from a non-negative stochastic process $\Lambda(\cdot).$ The goal is to approximate the intensity of a Cox process using a finite dimensional Gaussian Process (GP), $\Lambda_m(\cdot)$, subject to some constraints. \cite{pmlr-lopez-gauss-cox} propose a Bayesian paradigm for this, where the posterior distribution is supported on a constraint space, and further propose a Metropolis-Hastings algorithm  using a Gaussian proposal, truncated to this support. The truncated Gaussian leads to an intractable proposal, and the authors resort to approximate inference by estimating the intractable constants. 

Consider the example of \cite{adams-cox,pmlr-lopez-gauss-cox}, where the true intensity is
\[
\lambda(x) = 2\exp\left\{-\dfrac{x}{15} \right\} + \exp\left\{-\left(\dfrac{x-25}{10} \right)^2 \right\}\,,
\]
on the domain $S = [0, 50]$. For the finite-dimensional GP, they consider a set of equispaced knots $t_1, \dots t_m \in S$ such that $t_j = (j-1)/\Delta_m, $ where $\Delta_m = 50$, the Lebesgue measure on the set $S$. Let $\Lambda_m(\cdot)$ be the finite dimensional approximation of the GP $\Lambda(\cdot)$ constructed via piecewise linear interpolations at the knots $t_1, \dots t_m$:
\[
\Lambda_m(x) = \sum_{j=1}^m \phi_j(x)\xi_j\,,
\]
where the functions $\phi_j$ : $j = 1, \dots m$ are given by 
\[
\phi_j(x) := 
\begin{cases} 
1 - \left|\frac{x - t_j}{\Delta_m}\right|, & \text{if } \left|\frac{x - t_j}{\Delta_m}\right| \leq 1, \\
0, & \text{otherwise.}
\end{cases}
\]
Our aim is to estimate $\Lambda_m$ under the constraint that it belongs to a set of convex functions satisfying some constraints. As in \cite{pmlr-lopez-gauss-cox}, taking the non-negativeness constraint, this is equivalent to estimating $\xi = (\xi_1 \dots \xi_m)^\top$ with $\xi \in \mathcal{C}_+$, where $\mathcal{C}_+$ is a convex set defined by
\[
\mathcal{C}_+ = \{a \in \mathbb{R}^m : \text{ for all } j = 1, \dots m : a_j > 0\}\,.
\]

Let the $N_0$ independent observations from the inhomogeneous Poisson process be denoted by $(n_{\nu}, \Tilde{x}_{\nu})$, where $\nu = 1, 2, \dots N_0$. For the $\nu$th observation, $n_{\nu}$ denotes the number of instances observed and the vector $\Tilde{x}_{\nu} = (x_{1\nu}, x_{2\nu}, \dots, x_{n_\nu \nu})$ are iid realizations from the intensity function. 
Under the finite-dimensional approximation framework, the likelihood for the $\nu^{\text{th}}$ observation is then
\begin{align*}
    L(\xi | n_{\nu}, x_{1\nu}, \dots x_{n_\nu \nu})
    &\propto e^{-\sum_{j=1}^{m}c_j \xi_j } \prod_{i=1}^{n_\nu}\sum_{j=1}^m \phi_j(x_{i\nu})\xi_j\,.
\end{align*}
Here, $c_1 = c_m = \Delta_m/2$ and $c_j = \Delta_m$ for $1 < j < m$. \cite{pmlr-lopez-gauss-cox} assume a truncated Gaussian prior, truncated to $\xi \in \mathcal{C}_+,$ with mean $0$ and covariance matrix $\Gamma$, where $\Gamma$ is the matrix discretization of a squared exponential kernel: 
\[
\Gamma_{ij} = \sigma^2 \exp\left(-\frac{(t_i - t_j)^2}{2l^2} \right)\,,
\]
where we estimate $\sigma^2$ and $l$ using the defaults in the \texttt{R} package \cite{lopez2018lineqgpr}. Thus, the prior density is,
\[
f({\xi}) = \frac{\exp\{-\frac{1}{2}\xi^T\Gamma^{-1}\xi\}}{\int_{\mathcal{C}_+}\exp\{-\frac{1}{2}s^T\Gamma^{-1}s\}ds}, \quad \xi \in \mathcal{C}_+.
\]
% The unnormalized posterior is then given by
% \begin{align*}
% &\pi \left(\xi|(n_1, \tilde{x}_1), \dots ,(n_{N_0}, \tilde{x}_{N_0}) \right) \\
% &\propto  f(\xi) \prod_{\nu = 1}^{N_0} L(\xi | n_{\nu}, x_{1\nu}, \dots x_{n\nu}) \\ 
% & = \exp\left( -\frac{1}{2} \xi^\top \Gamma^{-1} \xi - c^\top \xi \right) \prod_{\nu = 1}^{N_0}\prod_{i=1}^{n} \phi^\top(x_{i\nu})\xi\,,
% \end{align*}
% % 
% where $c = [c_1, \dots c_m]^T$ and $\phi(x) = [\phi_1(x) \dots \phi_m(x)]^T.$ 

In order to sample from the $m$-dimensional posterior distribution, the authors employ a Metropolis-Hastings sampler with a truncated multivariate Gaussian proposal for an $m \times m$ matrix $\Sigma$ having density
\[
q(\chi|\xi_k) = \frac{
\exp\left( -\frac{1}{2} [\chi - \xi_k]^\top \Sigma^{-1} [\chi - \xi_k] \right)
\mathbb{I}(\chi \in \mathcal{C}_+)}{
\int_{\mathcal{C}_+} \exp\left( -\frac{1}{2} [t - \xi_k]^\top \Sigma^{-1} [t - \xi_k] \right) dt
}
.
\]
This choice is particularly critical since the posterior distribution is bound to have high correlation across adjacent components. The use of a general $\Sigma$ in the proposal allows for preconditioning on the truncated space, but also renders the proposal density intractable. We employ a warm-up run of an MCMC implementation to estimate the covariance structure of the proposal and set $\Sigma$ to be $\eta$ times this estimated sample covariance. 
% The scaling $\eta$ is tuned for desired acceptance probability. 

Let $s( \cdot | \xi_k)$ denote the density of an untruncated multivariate Gaussian
% \begin{align*}
% s(\chi|\xi_k) = \frac{1}{(2\pi|\Sigma|)^{n/2}}e^{ -\frac{1}{2} (\chi - \xi_k)^\top \Sigma^{-1} (\chi - \xi_k) }, \qquad {\chi \in \mathbb{R}^m}\,,
% \end{align*}
and define the following integral
\[
r(\xi_{k}) = \int_{ \mathcal{C}_+} s(t|\xi_{k}) dt.
\]
% The covariance matrix of the proposal $\Sigma$ is given by 
% \[
% \Sigma = \eta \Gamma\,,
% \]
% where $\Gamma$ is as defined above and $\eta$ is tuned for a desired acceptance probability. 
Since the proposal density is intractable, \cite{pmlr-lopez-gauss-cox} approximate the normalizing constants $r(\xi_k)$ and $r(\chi)$ in each iteration of the Metropolis-Hastings using Monte-Carlo techniques suggested by \cite{Botev_2016} with 200 Monte Carlo samples for each integral evaluation. This turns out to be a relatively expensive procedure, in addition to it being fairly inexact, as we will demonstrate. 

We now explain the details of implementing the two-coin algorithm. For any step of the Markov chain, let the current iterate be $\xi_{k}$ and the proposed value be $\chi$. The Barker's acceptance ratio is
% \begin{align*}
%     \alpha_B =\\
%     & \frac{\Tilde{f}_{\xi|\Tilde{x}}(\xi_{k+1})s(\xi_k | \xi_{k+1})}{\Tilde{f}_{\xi|\Tilde{x}}(\xi_{k+1})s(\xi_k | \xi_{k+1}) + \Tilde{f}_{\xi|\Tilde{x}}(\xi_{k})s(\xi_{k+1} | \xi_{k})}
% \end{align*}
% where $\tilde{f}$ is the unnormalized posterior. This ratio simplifies to:
\begin{align*}
    \alpha_B(\chi, \xi_k) =
    &\frac{\pi(\chi) r(\xi_k)}{\pi(\chi) r(\xi_k) + \pi(\xi_k) r(\chi)}\,.
\end{align*}
Notice that,
\begin{align*}
    r(\xi_k) &= \int_{t \in \mathcal{C}_+} s(t| \xi_{k}) \leq 1 =: b_x\,.
\end{align*}
Analogously,
\[
r(\chi) = \int_{t \in \mathcal{C}_+} s(t| \chi) \leq 1 =: b_y\,.
\]
The corresponding values of $p_x$ and $p_y$ are
\[
p_x = \int_{t\in \mathcal{C}_+}s(t|\chi)dt \quad \text{and} \quad p_y =\int_{t\in \mathcal{C}_+}s(t|\xi_k)dt.
\]
To draw a Bern$(p_y = r(\xi_k)/b_x)$ we do the following, setting $f(\cdot) = s(\cdot | \xi_k)$:
\begin{enumerate}
    \item Draw $M \sim s(\cdot|\xi_{k})$.
    % \item If $M \in \mathcal{C}_+$, set $p_y = 1$, else, set $p_y = 0$.
    \item Set $C_2|M = \mathbb{I}(M \in \mathcal{C}_+)$. Then $C_2 \sim \text{Bern}(p_y)$.
\end{enumerate}

\begin{figure}
\centering
\includegraphics[width=0.55\columnwidth]{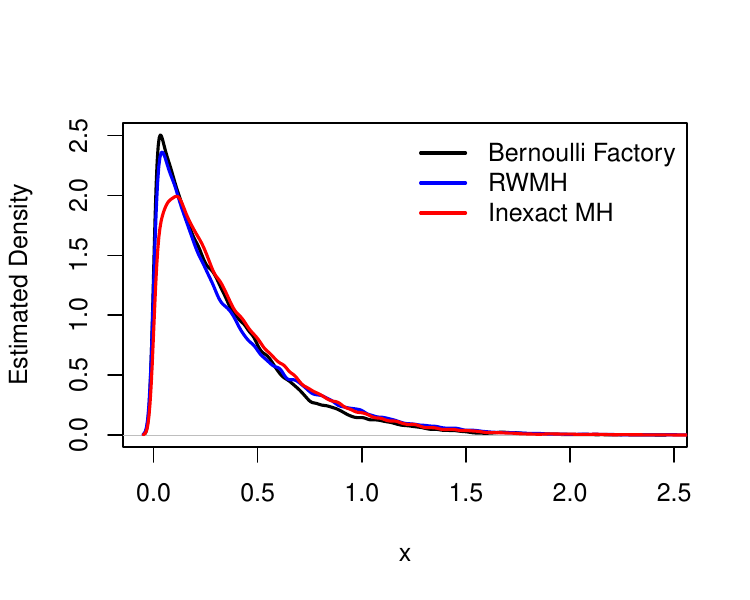}
\caption{Cox: Estimated density plots for the 100th component using all three methods.}
\label{fig:cox-comp}
\end{figure}

We generate the data with $N_0 = 10$ comparing the performance of the exact Bernoulli factory Barker's algorithm with the inexact Metropolis-Hastings implemented by \cite{pmlr-lopez-gauss-cox}, for this same intensity function. We also run a random-walk Metropolis Hastings algorithm, with the non-truncated proposal. Similar to \cite{pmlr-lopez-gauss-cox}, we set $m = 100$. For Metropolis-Hastings, we tune $\eta$ to obtain $23\%$ acceptance and for Barker's we tune $\eta$ to obtain $16\%$ acceptance.  In Figure~\ref{fig:cox-comp} we show the estimated marginal density plot for the $100$th component of $\xi$ from all the methods using Markov chains of length $10^6$. There is a noticeable difference in the  estimated density plots of the exact and the inexact methods.
% ; with Barker's being exact, it is then evident that the inexact implementation is visibly inaccurate.

Replicating the Markov chains of length $2 \times 10^5$, 100 times, the results in Table~\ref{tab:multiess_comparison} show that the Barker's implementation is actually faster, enough to offset the reduction in effective sample size, with the added advantage of exact inference. The reason the two-coin algorithm is fast is because of relatively low number of Bernoulli factory steps required; the average number of Bernoulli factory loops required in a $2 \times 10^5$ run was $18.39$, with the average maximum number of loops being $2427$.
\begin{table}
\centering
% \small 
% \tabcolsep=6pt
\caption{Cox: ESS and computational time}
\label{tab:multiess_comparison}
\begin{tabular}{lccc}
\hline
  & \textbf{Barker's exact} & \textbf{Inexact MH} & \textbf{Random Walk MH}\\
\hline
ESS &  1108  &  1531 & 473 \\
ESS$/$s  & 0.3906  & 0.0709  & 0.2485\\
Time (s) & 2950 & 21988 & 1984 \\
\hline
\end{tabular}
\end{table}

\section{Discussion}

The two-coin algorithm proves to be significantly more efficient when utilized for intractable proposals, than when dealing with doubly intractable posteriors. This is almost exclusively because the proposals must be such that sampling from them is viable. This limits the general complexity of the proposal density making them inherently convenient to bound. As shown in our examples, natural bounds are immediately available when dealing with truncated proposals. Certainly, it is possible that for some implementations the bounds are weak and the two-coin algorithm is inefficient. In such a case, \cite{vats2021efficient} propose a portkey two-coin algorithm that can be significantly more efficient than the two-coin algorithm. However, this Bernoulli factory induces a different acceptance probability, which although still yielding $\pi$-invariance, is statistically more inefficient than Barker's algorithm.

%%%%%%%%%%%%%%%%%%%%%%%%%%%%%%%%%%%%%%%%%%%%%%
%% Support information, if any,             %%
%% should be provided in the                %%
%% Acknowledgements section.                %%
%%%%%%%%%%%%%%%%%%%%%%%%%%%%%%%%%%%%%%%%%%%%%%
% \begin{acks}[Acknowledgments]
% The authors would like to thank the anonymous referees, an Associate
% Editor and the Editor for their constructive comments that improved the
% quality of this paper.
% \end{acks}

% %%%%%%%%%%%%%%%%%%%%%%%%%%%%%%%%%%%%%%%%%%%%%%
% %% Funding information, if any,             %%
% %% should be provided in the                %%
% %% funding section.                         %%
% %%%%%%%%%%%%%%%%%%%%%%%%%%%%%%%%%%%%%%%%%%%%%%

\section{Acknowledgments}
Dootika Vats is supported by Google Asia Pacific Pte Ltd.

\bibliographystyle{apalike} 
\bibliography{bibliography}       

\end{document}